\documentclass[pra,twocolumn,a4paper,groupedaddress,showpacs,floatfix,aps,10pt,
longbibliography,nobibnotes]{revtex4-1}

\usepackage{graphicx,amsmath,amssymb,amsfonts,dsfont,subfigure,color}

\begin{document}

\title{Collimated UV light generation by two-photon excitation to \\ a Rydberg state in Rb vapor}

\author{Mark Lam${}^{1}$}
\author{Sambit B. Pal${}^{1}$}
\author{Thibault Vogt${}^{1,2}$}
\author{Christian Gross${}^{1}$}
\author{Martin Kiffner${}^{1,3}$}
\author{Wenhui Li${}^{1,4}$}

\affiliation{Centre for Quantum Technologies, National University of Singapore, 3 Science Drive 2, Singapore 117543${}^1$}
\affiliation{MajuLab, CNRS-UNS-NUS-NTU International Joint Research Unit UMI 3654, Singapore 117543${}^2$}
\affiliation{Clarendon Laboratory, University of Oxford, Parks Road, Oxford OX1 3PU, United Kingdom${}^3$}
\affiliation{Department of Physics, National University of Singapore, Singapore 117542${}^4$}




\begin{abstract}

We use two continuous-wave (CW) laser beams of 780 nm and 515 nm to optically drive $^{85}$Rb atoms in a heated vapor cell to a low-lying Rydberg state 10D$_{5/2}$. We observe a collimated ultraviolet (UV) beam at 311 nm, corresponding to the transition frequency from the 11P$_{3/2}$ state to the 5S$_{1/2}$ state. This indicates the presence of a coherent four-wave mixing process, built up by two input laser fields as well as a terahertz (THz) radiation of 3.28 THz that is generated by amplified spontaneous emission between the 10D$_{5/2}$ and the 11P$_{3/2}$ states. We characterize the 311 nm UV light generation and its dependence on various physical parameters. This scheme could open up a new possibility for generating narrow-band THz waves as well as deep UV radiation.

\end{abstract}


\maketitle

\smallskip


Laser-induced coherence among quantum states in an atomic medium leads to greatly enhanced nonlinear susceptibilities compared to that of solid state crystal media~\cite{coherent_nonlinear}, and gives rise to fascinating phenomena such as electromagnetically induced transparency (EIT)~\cite{fleischhauer2005electromagnetically}, coherent population trapping (CPT)~\cite{arimondo1996v}, and lasing without inversion (LWI)~\cite{scully1989degenerate}. Nonlinear optical processes in such coherently prepared atomic media have been extensively studied for a wide range of applications. In particular, four-wave mixing (FWM) in alkali atomic gases is used to produce entangled states of light~\cite{Pooser:09} and photon pairs~\cite{srivathsan2013narrow}, to construct quantum memories~\cite{radnaev2010quantum}, and to transfer orbital angular momentum between near-infrared and blue light fields~\cite{walker2012trans}. A very active area of research on FWM has been the investigation of frequency up and down conversion by using continuous wave excitation lasers in hot atomic vapors, which enables the efficient generation of new coherent optical fields in the visible and infrared regions~\cite{zibrov2002efficient, Meijer:06, Akulshin:09, Vernier:10, Brekke:13, Akulshin:14, Sell:14, Offer:16, becerra2008nondegenerate, schultz2009coherent,koelle2012,demelo2014}.

\begin{figure}[tbp]
\centering
\includegraphics[width= 0.9 \linewidth]{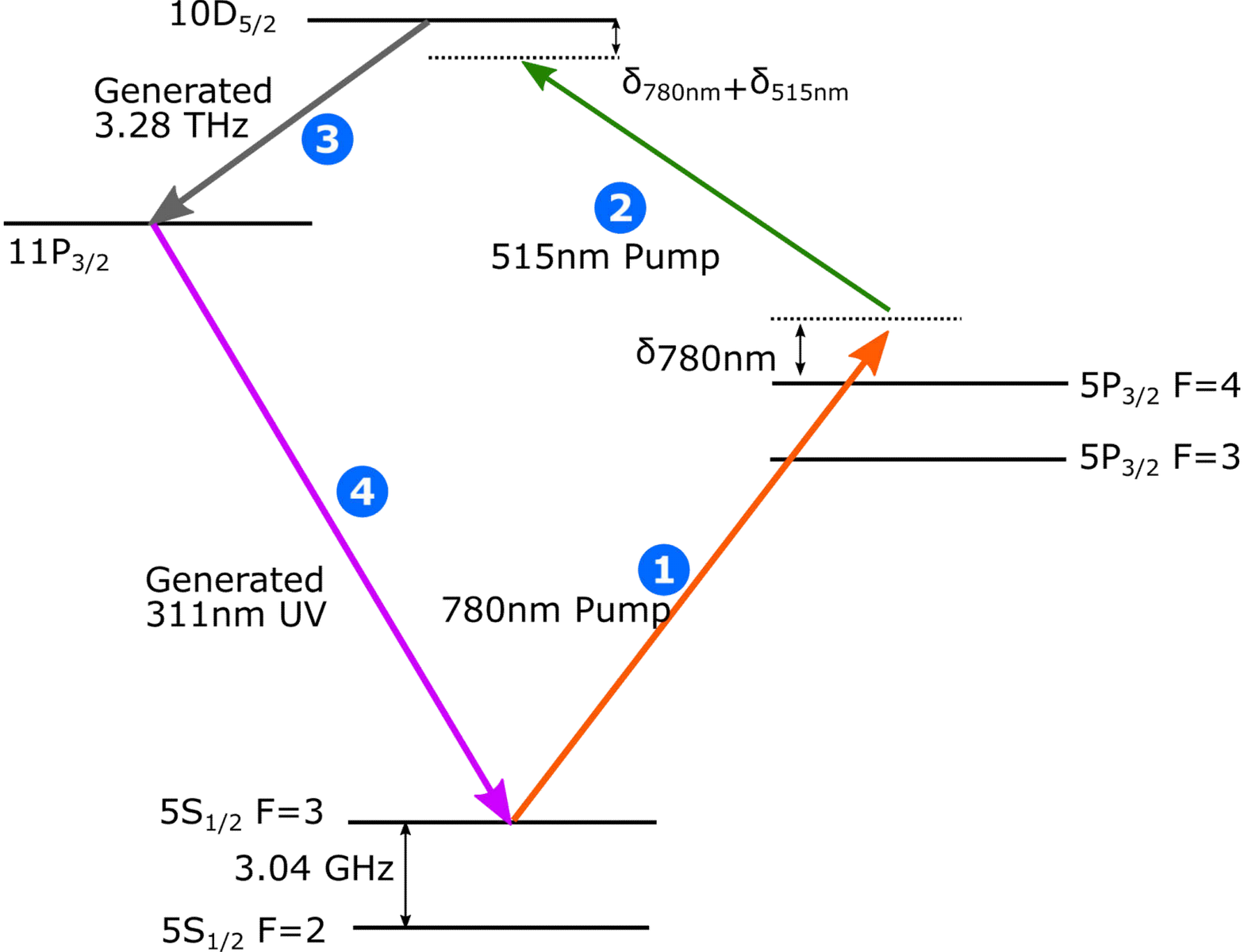}
\caption{Energy levels of $^{85}$Rb relevant to our scheme. The detuning of the 780 nm pump laser from the 5S$_{1/2}$ F=3 $\rightarrow$ 5P$_{3/2}$ F=4 transition resonance is denoted as $\delta_\textrm{780nm}$, while $\delta_\textrm{780nm} + \delta_\textrm{515nm}$ is the two-photon detuning from the resonance frequency between the 5S$_{1/2}$ F=3 and 10D$_{5/2}$ states. Numeric labels have been assigned to each transition for the reader's convenience.}
\label{scheme}
\end{figure}


In this paper, we investigate the generation of collimated UV light using $^{85}$Rb in a thermal natural rubidium atomic vapour via a parametric four-wave mixing process involving the low-lying Rydberg states 10D$_{5/2}$ and 11P$_{3/2}$. The relevant energy levels and the corresponding transitions of the experiment are shown in Fig.~\ref{scheme}. The two transitions, 5S$_{1/2}$ $\rightarrow$ 5P$_{3/2}$ (transition 1) and 5P$_{3/2}$ $\rightarrow$ 10D$_{5/2}$ (transition 2), are driven by two near-resonant laser fields at 780~nm and 515~nm, respectively. We experimentally observe a collimated and single wavelength laser beam at 311 nm, which corresponds to the 11P$_{3/2} \rightarrow$ 5S$_{1/2}$ transition (transition 4 in Fig.~\ref{scheme}). We attribute this generated 311 nm beam to a four-wave mixing process from the two input lasers as well as the 3.28 THz radiation emitted on the 10D$_{5/2} \rightarrow$ 11P$_{3/2}$ transition (transition 3 in Fig.~\ref{scheme}). The generation of this THz radiation via amplified spontaneous emission is facilitated by the long radiative lifetime of the 10D$_{5/2}$ state ($\sim$780 ns)~\cite{beterov2009}.

Note that a similar scheme involving the lower lying states 5D$_{5/2}$ and 6P$_{3/2}$ has been studied in several experiments using rubidium vapors ~\cite{zibrov2002efficient, Meijer:06, Akulshin:09, Vernier:10, Brekke:13, Akulshin:14, Sell:14, Offer:16}, where a collimated blue light at 421 nm, corresponding to the transition 6P$_{3/2} \rightarrow$ 5S$_{1/2}$, is produced via a four-wave mixing process. A generated 5.2 $\mu$m infrared radiation corresponding to the transition 5D$_{5/2} \rightarrow$  6P$_{3/2}$,  enabling the FWM process, has been observed in some of the experiments \cite{Akulshin:14}. The difference between our scheme and the works in ~\cite{zibrov2002efficient, Meijer:06, Akulshin:09, Vernier:10, Brekke:13, Akulshin:14, Sell:14, Offer:16} is that, instead of the 5D$_{5/2}$ state, a low-lying Rydberg state is being excited by pump lasers. As a result, two very different frequencies outside the typical optical regime are generated, namely a THz field on transition 3 and a UV field on transition 4. In our current setup, the direct measurement of the generated 3.28 THz field is hindered by the opacity of the fused-silica vapour cell windows at THz frequencies. However, the presence of the THz field is firmly established by the observation of the collimated 311 nm UV light.


\begin{figure}[tb]
	\centering
	\includegraphics[width=0.9\linewidth]{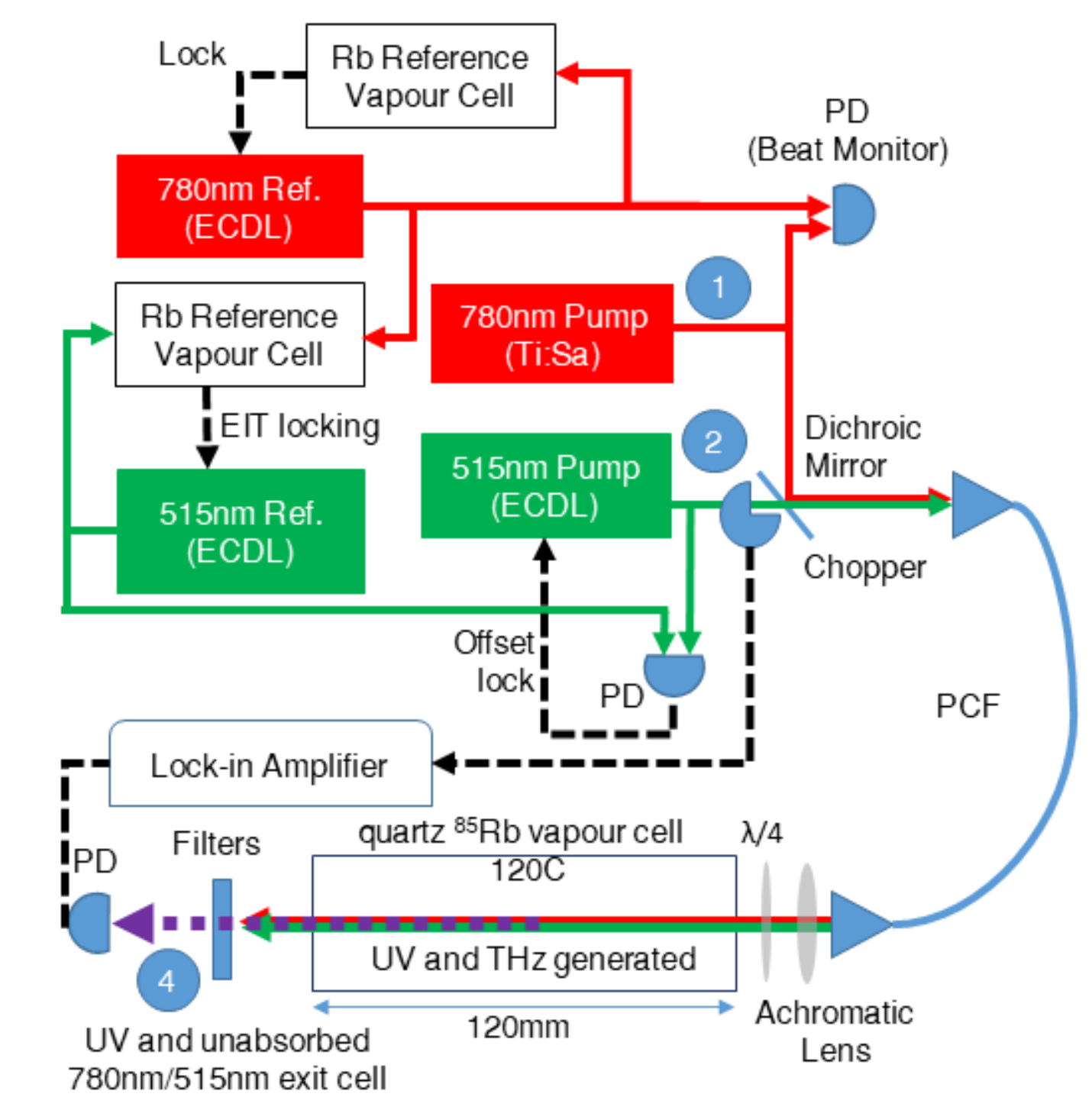}
	\caption{Schematic representation of the experimental setup. The various optical paths have been assigned numeric labels matching their associated transitions as per Fig.~\ref{scheme}. The cell temperature is regulated to $\sim$120 $^{\circ}$C for most measurements presented here. Details are given in the text.}
	\label{setup}
\end{figure}

Our experimental setup is illustrated in Fig.~\ref{setup}. Two pump lasers consist of a 780 nm Titanium Sapphire (Ti:Sa) laser driving transition 1 and a 515 nm external cavity diode laser (ECDL) driving transition 2. The frequency of the 780 nm pump laser is determined via an optical beat note measurement with a 780 nm reference laser, which is locked on the 5S$_{1/2}$ F=3 to 5P$_{3/2}$ F=4 transition. Meanwhile, the 515 nm pump laser is stabilized to a 515 nm reference laser via frequency offset locking~\cite{beatlock}. We lock this 515 nm reference laser to the spectrum of a ladder-type EIT transition formed with the 10D$_{5/2}$, 5P$_{3/2}$ and 5S$_{1/2}$ states (transitions 1 and 2). The EIT is realized in our setup by overlapping the two reference lasers, both right-circularly polarized, in a counter-propagating configuration through a Rb vapour cell~\cite{mohapatra2007coherent,csadamseitlock}.

The two pump beams are combined on a dichroic mirror and coupled into a photonic crystal fiber (PCF) to ensure a perfect spatial overlap. The output from the PCF is configured for right circular polarization and focused with a single achromatic lens, and then sent into a commercially-available UV fused silica vapour cell containing an isotopic mixture of rubidium per its natural abundance. The beam waists are located near the center of the cell, with $1/e^2$ diameters of $\sim$210 $\mu$m. The cell is housed in a temperature-stabilized copper assembly using heatpipes with bifilar windings of heating wire. This increases the coil distance from the cell location and consequently reduces the effect of stray magnetic fields. At the exit end of the cell, we block the transmitted input pump lights with the use of UV-bandpass interference filters, and measure the power of the generated light using a calibrated UV-enhanced photodiode (PD).


\begin{figure}[tb]
\centering
\includegraphics[width=0.5\linewidth]{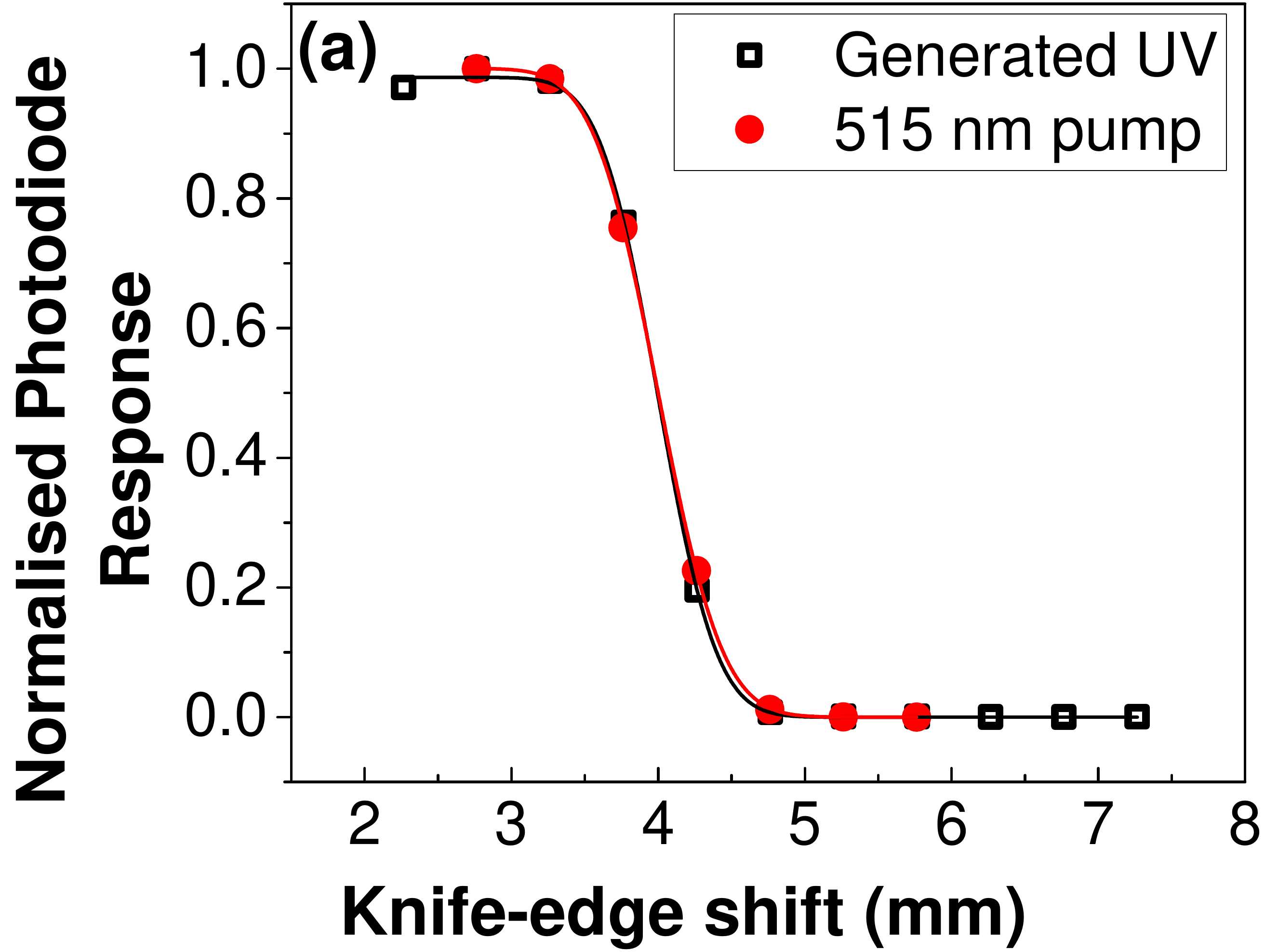}
\includegraphics[width=0.45\linewidth]{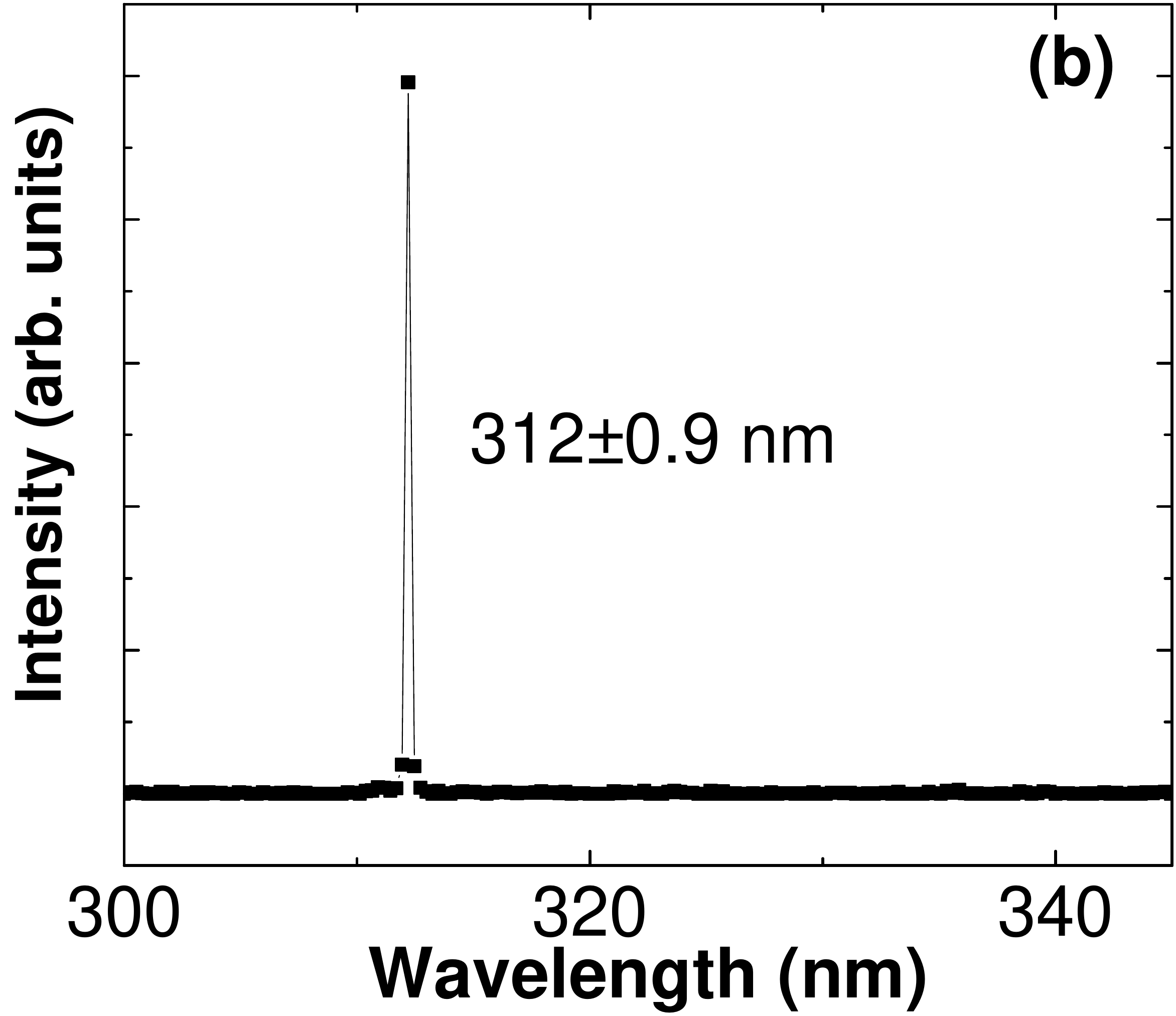}
\caption{(a) Knife-edge measurements of the generated UV light and the 515 nm pump light at 19 cm from the center of the vapour cell. The solid lines are the fits to the data with the error function, which is the integration of a Gaussian beam profile along the radial direction. (b) The spectrum of the output beam measured by a narrow-range grating spectrometer.}
\label{mightex}
\end{figure}

We observe a UV beam at the exit of the vapor cell co-propagating in the same direction as the input pump beams, over a range of experimental conditions. We first characterize the spatial profile of this UV beam by using the well-known knife-edge method~\cite{0022-3735-5-3-015, Khosrofian:83}. The measurement results in Fig.~\ref{mightex}(a) show that the generated UV light and the input 515 nm beam overlap with each other, and have a Gaussian profile with nearly the same beam size. This directionality of the generated UV light is consistent with the phase-matching condition for four-wave mixing $\mathbf{k}_1 + \mathbf{k}_2 = \mathbf{k}_3 + \mathbf{k}_4$, where $\mathbf{k}_i$ is the wave vector corresponding to transition \textit{i} (\textit{i}$\in$ \{1,2,3,4\}).

\begin{table*}[!htbp]
\centering
\linespread{0.7}
\rmfamily
\begin{tabular}{c |c c c |c c c | c}
\hline
   & \multicolumn{3}{c|}{10D$_{5/2} \rightarrow n$P$_{3/2}$}         & \multicolumn{3}{c|} {$n$P$_{3/2} \rightarrow$ 5S$_{1/2}$}  &                           \\ \hline
n  & $\nu$ (THz)     & $\lambda$ (nm)  & $|d_{\mathrm{DP}}|$ (ea$_0$)          & $\nu$ (THz)     & $\lambda$ (nm)  & $|d_{\mathrm{PS}}|$ (ea$_0$) &  $|d_{\mathrm{DP}}\cdot d_{\mathrm{PS}}|^2$ ((ea$_0$)$^4$)      \\ \hline
11 & 3.3             & 91221.5         & 63.64                      & 962.9           & 311.3           & 0.03              &  3.645                          \\
10 & 17.0            & 17637.1         & 1.44                       & 949.2           & 315.8           & 0.04             &  0.0033                         \\
9  & 37.7            & 7950.0          & 1.24                      & 928.5           & 322.9           & 0.05              &  0.0038                           \\
8  & 71.2            & 4212.3          & 0.83                       & 895.0           & 335.0           & 0.05             &  0.0017                          \\
7  & 130.6           & 2294.6          & 0.55                      & 835.5           & 358.8           & 0.10             &  0.0030                           \\
6  & 252.9           & 1185.5          & 0.36                       & 713.3           & 420.3           & 0.27             &  0.0095                           \\
5  & 581.9           & 515.2           & 0.22                      & 384.2           & 780.2           & 2.98             &  0.4298                           \\ \hline
\end{tabular}
\caption{Table of frequencies ($\nu$), wavelengths ($\lambda$), and dipole matrix elements ($d_{DP}$ and $d_{PS}$) for the transitions 10D$_{5/2} \rightarrow n$P$_{3/2}$ and $n $P$_{3/2} \rightarrow$ 5S$_{1/2}$~\cite{ARC2017}. Given in the last column are the values of $|d_{\mathrm{DP}} \cdot d_{\mathrm{PS}}|^2$. The 10D$_{5/2} \rightarrow n'$F transitions and further cascade transitions from $n'$F states back to the 5S$_{1/2}$ state are not considered here.} \label{dipolematrix}
\end{table*}

The wavelength of the generated UV beam is measured with a grating spectrometer, and the only detectable peak is at $\sim$311 nm, as shown in Fig.~\ref{mightex} (b). This wavelength corresponds to the 11P$_{3/2}$ $\rightarrow$ 5S$_{1/2}$ transition. None of the other UV wavelengths listed in Table~\ref{dipolematrix}, corresponding to the transitions $n$P$_{3/2} \rightarrow$ 5S$_{1/2}$ with $n < $11, are visible on this power scale. We make the following simple arguments to explain this observation. The third order nonlinear susceptibility of FWM in atomic media is proportional to the product of the dipole matrix elements of the four corresponding transitions~\cite{fleischhauer2005electromagnetically}. Since all the possible FWM pathways in Table~\ref{dipolematrix} share the same two pumping transitions, the generated UV power of a FWM pathway, proportional to the square of the third order nonlinearity, will largely be determined by $|d_{\mathrm{DP}} \cdot d_{\mathrm{PS}}|^2$, where $d_{\mathrm{DP}}$ and $d_{P\mathrm{S}}$ are the dipole moments of the two cascade transitions. The values given in the last column of Table~\ref{dipolematrix} indicate that the nonlinearity of the four-wave mixing via transitions 3 and 4 is much larger than that of other FWM pathways, which results in the domination of the 311 nm in the generated light.

\begin{figure}[b]
	\centering
	\includegraphics[width=0.9\linewidth]{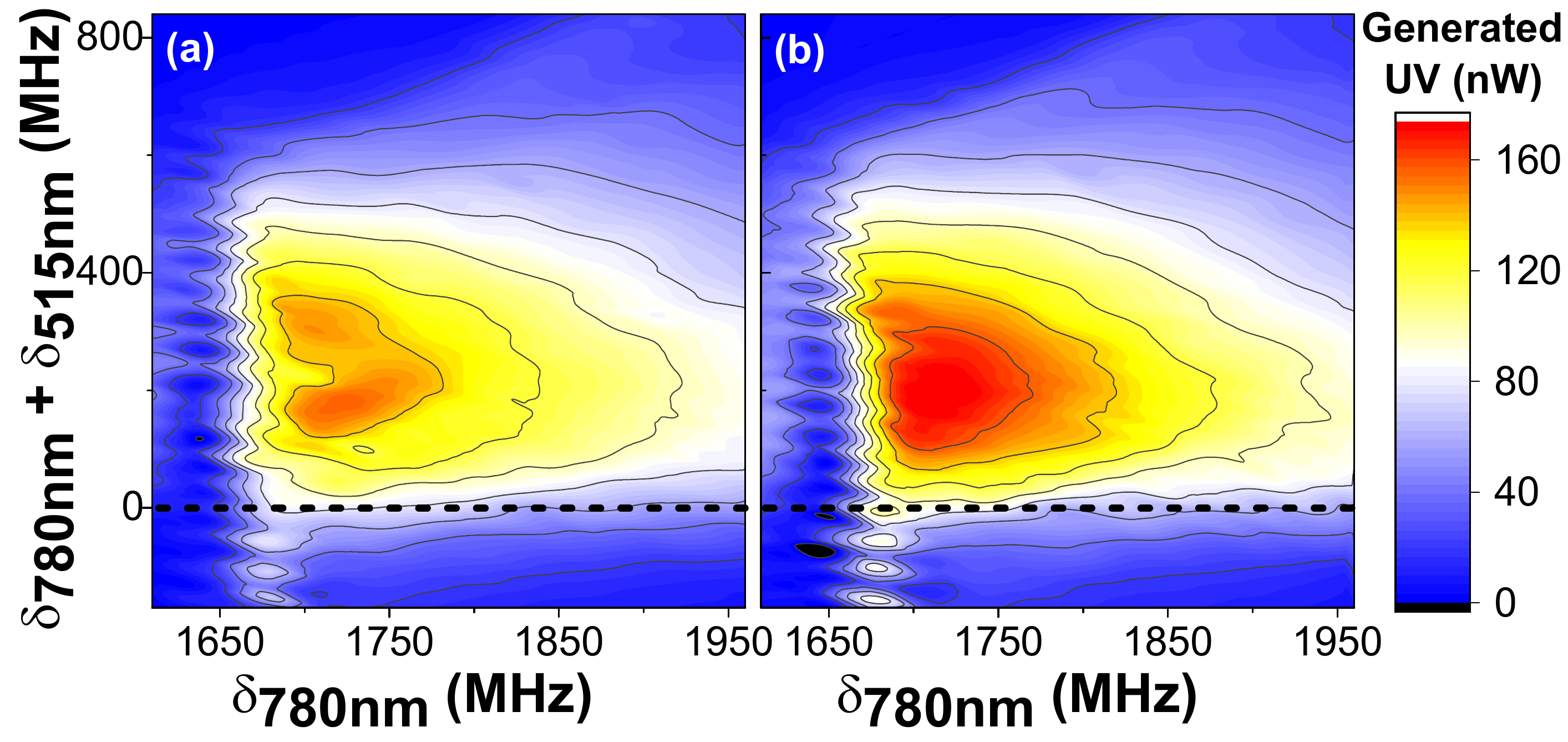}
	\caption{Generated UV power as a function of the 780 nm pump laser detuning ($\delta_{780nm}$) and the two-photon detuning ($\delta_{780nm}$+$\delta_{515nm}$) without the repump laser (left) and with the repump laser (right). The horizontal dashed line denotes the atomic two-photon resonance $\delta_{780nm}$+$\delta_{515nm}$ = 0. The cell temperature is 126$^\circ$C, while the 515 nm and 780 nm pump lasers have a power of 19 mW and 150 mW, respectively.}

		\label{detuning}
\end{figure}

The above spatial and spectral measurements confirm that this collimated UV beam is generated by four-wave mixing instead of spontaneous emission. Next we discuss the dependence of the UV generation on the two pump laser frequencies by scanning them independently. In Fig.~\ref{detuning} (a), the generated UV power is plotted as a function of the 780 nm detuning ($\delta_{780nm}$) and the two-photon detuning ($\delta_{780nm} + \delta_{515nm}$). The optimal UV generation is observed near the two-photon resonance of transitions 1 and 2 with a width of around 250 MHz, which we attribute to Doppler broadening. Along the horizontal axis, an asymmetric peak appears with a sharp edge at around $\delta_{780nm}$ = 1.65 GHz and a long tail of width $\approx$ 800 MHz for $\delta_{780nm} >$ 1.65 GHz. This feature can be understood as follows. We have experimentally verified that the strong Kerr-lensing near the resonance of transition 1 distorts the 780 nm pump beam, and renders the FWM process ineffective. The beam restores its original shape at a detuning $\delta_{780nm} \sim$ 1.6 GHz, where the coherence of FWM survives and produces the 311 nm light that we observe. Note that a similar effect due to Kerr-lensing has been observed before in the generation of 421 nm light via FWM \cite{Vernier:10}.

To check whether the optimal detuning of $\delta_{780nm} \sim$1.6 GHz is related to the coherence induced by coupling to both hyperfine levels of the 5S$_{1/2}$ state \cite{Meijer:06}, we repeat the spectroscopic measurements with an additional repumper ECDL locked on the 5S$_{1/2}$ F=2 $\rightarrow$ 5P$_{1/2}$ F=2 transition of the D1 line. This repumper laser will redistribute most of the ground state population into the hyperfine level 5S$_{1/2}$ F=3. Fig.~\ref{detuning} (b) shows that this repumper causes no shift in the optimal detuning compared to that in Fig.~\ref{detuning} (a), but results in an overall increase of 15\% in the UV output. This result suggests that the UV generation observed is mostly due to the population in the 5S$_{1/2}$ F=3 level.

\begin{figure}[b]
	\centering
	\includegraphics[width=0.8\linewidth]{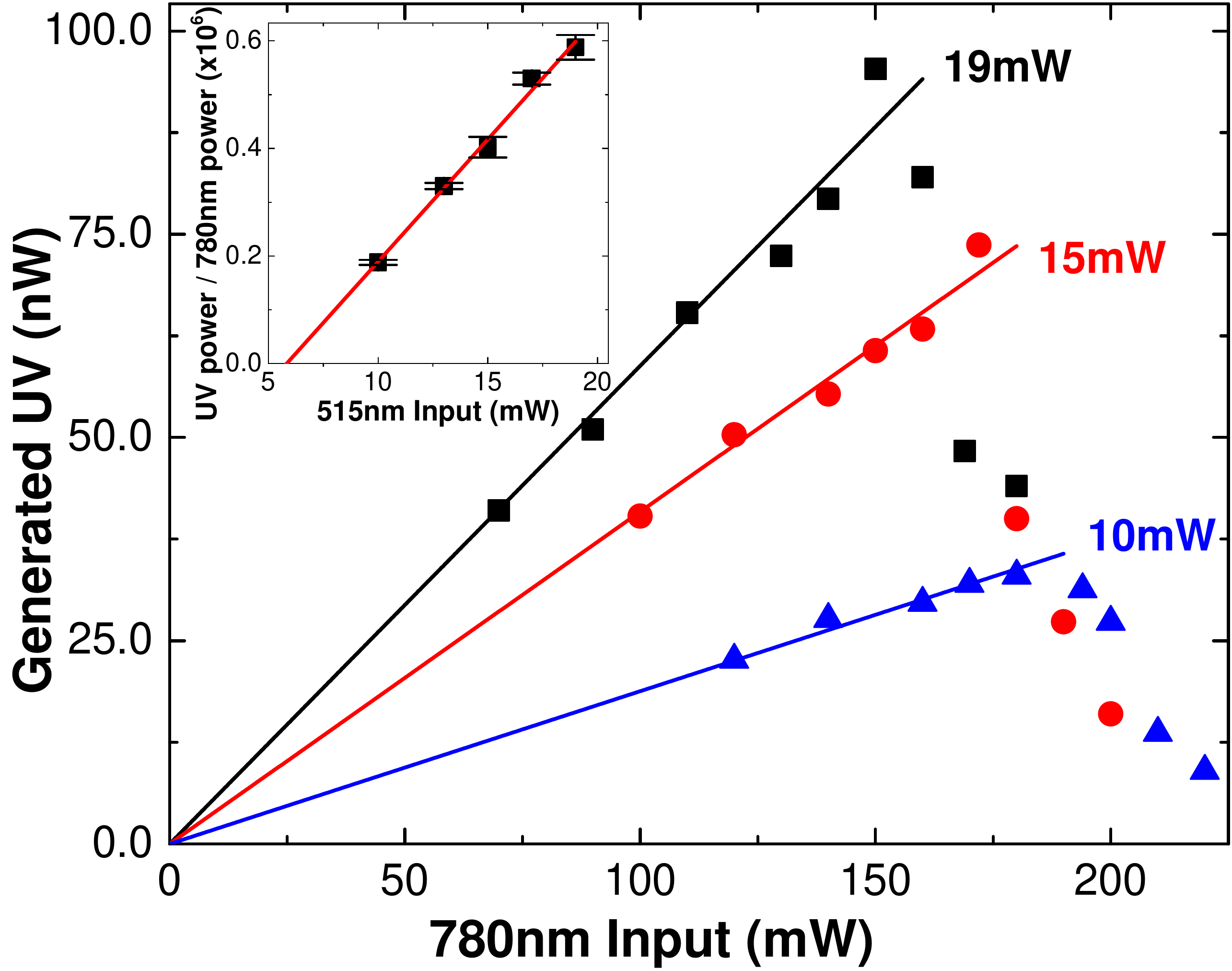}
	\caption{Generated UV power as a function of the 780 nm pump power for the 515 nm pump powers of 19 mW (black square), 15 mW (red circle), and 10 mW (blue triangle). The data is taken at 123$^\circ$C. The solid lines are linear fits to the data of lower 780 nm powers, and the slopes of these fits with fitting errors are plotted as a function of the 515 nm power in the inset. The solid line in the inset is a linear fit.}
	\label{powdep}
\end{figure}

We investigate the dependence of the generated UV light on the input laser powers, while keeping the detuning of both lasers at the values where maximum UV generation is observed. In Fig.~\ref{powdep}, all curves show that the generated UV power first increases with increasing 780 nm power. A further increase of the 780 nm power results in a reduction of the UV generation. To better understand this dependence, we simulate the light propagation inside the vapor cell within a simple four-level model and semi-classical Maxwell-Bloch equations~\cite{han2018} where atomic motion is accounted for by transit relaxation~\cite{sagle1996}. We find that the predicted UV power and its dependence on the pump powers agree qualitatively with the measurements. In particular, our simulations suggest that the reduction of the generated UV light at large 780nm input powers can be explained by a reduction of the 5S $\rightarrow$ 5P $\rightarrow$ 10D excitation efficiency due to the light-induced dressing of the 5P states. In the inset of Fig.~\ref{powdep}, the slopes of the linear fits at small 780 nm powers are shown to increase linearly with the 515 nm pump power, and has no sign of saturation within the range of 515 nm power available in our experiment. This indicates that using higher 515 nm power will improve the UV generation.

\begin{figure}[tb]
	\centering
	\includegraphics[width= 0.9 \linewidth]{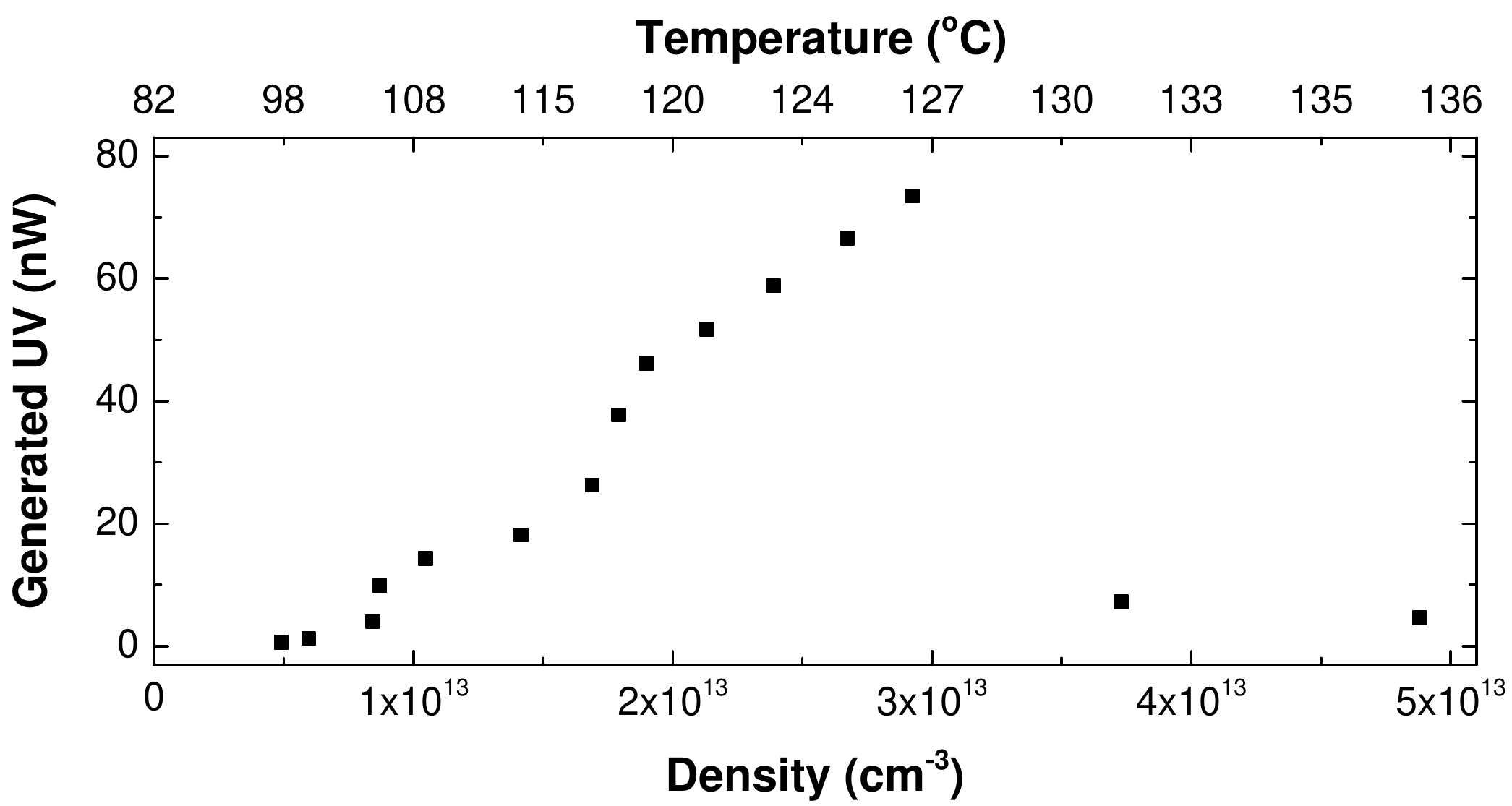}
	\caption{Generated UV power as a function of atomic density. The vapor cell temperature is controlled externally to change the atomic density, and is given along the horizontal axis at the top, while the atomic density of the corresponding saturated vapor pressure is given along the axis at the bottom.}
	\label{density}
\end{figure}

The density-dependence of the generated UV power is shown in Fig.~\ref{density}, where both detunings and 780nm pump power are configured for maximum UV output at a fixed 515 nm pump power of 14 mW. At low densities, the generated UV power increases steadily with atomic density. However, UV generation falls sharply to almost zero for densities larger than 3 $\times$ 10$^{13}$ cm$^{-3}$, which persists even with increasing 780 nm power and/or adjusting the laser detunings. This abrupt reduction may be due to ions created by collisions between Rydberg atoms~\cite{weller2016,wade2018}, which induce Stark shifts of atomic levels. Further investigation is needed to verify this assumption.


In conclusion, we have demonstrated the generation of deep UV light via FWM involving low lying Rydberg states. Despite multiple decay pathways from the upper Rydberg level 10D$_{5/2}$, the dominant FWM process involves the most strongly dipole-coupled Rydberg level 11P$_{3/2}$. This results in the generation of a nearly monochromatic UV beam, while other FWM pathways give rise to a much weaker nonlinearity and are thus strongly suppressed.

Besides being useful for up conversion to deep UV radiation, this scheme of optically pumping atomic media into low-lying Rydberg states may offer a possibility for building up a continuous wave (CW) THz source with high spectral brightness. This is especially interesting, given that there has been tremendous efforts \cite{Lewis_thzreview, thzroadmap, qcldevelopments} to develop compact, narrow-linewidth, high-power THz sources. Even though it is not possible to extract the THz radiation in our current apparatus, our theoretical simulations allow us to estimate the amount of THz radiation generated on the 10D$_{5/2} \rightarrow$ 11P$_{3/2}$  transition.  We find that, under the current experimental conditions, the maximal power of the THz beam co-propagating with the other beams is of the order of 1 - 10 $\mu$W  before the exit of the vapor cell. Note that in our system the vast majority of THz photons is generated via amplified spontaneous emission and not four-wave mixing due to the large imbalance between the dipole moments of transitions 3 and 4, $d_{DP}/d_{PS} >>$ 1 (see Table~\ref{dipolematrix}). To facilitate extraction of the THz radiation, a new vapour cell with silicon output window is being constructed. To further increase the power of the generated UV or THz fields, a build-up cavity may be used. In summary, we believe that our experimental demonstration lays the ground work for further utilizing this scheme to generate coherent deep UV and THz radiations not easily accessible by other methods.

\section*{Acknowledgments}

The authors acknowledge the support by the National Research Foundation, Prime Minister's Office, Singapore and the Ministry of Education, Singapore under the Research Centres of Excellence programme.



\end{document}